\documentclass[10pt,letterpaper]{article}
\usepackage{opex3}
\usepackage{amsmath}

\usepackage{graphicx}
\usepackage{dcolumn}
\usepackage{bm}
\usepackage{float}
\usepackage{subfig}
\usepackage{wrapfig}
\usepackage{epstopdf}

\begin{document}


\title{Probing the dynamics of an optically trapped particle by phase sensitive back focal plane interferometry}

\author{Basudev Roy, Sambit Bikas Pal, Arijit Haldar, Ratnesh Kumar Gupta, Nirmalya Ghosh, and Ayan Banerjee}

\address{Department of Physical Sciences, IISER-Kolkata}\thanks{}
 \email{ayan@iiserkol.ac.in}
\date{\today}

\begin{abstract}
The dynamics of an optically trapped particle are often determined by measuring intensity shifts of the back-scattered light from the particle using position sensitive detectors. We present a technique which measures the phase of the back-scattered light using balanced detection in an external Mach-Zender interferometer scheme where we separate out and beat the scattered light from the bead and that from the top surface of our trapping chamber. The technique has improved axial motion resolution over intensity-based detection, and can also be used to measure lateral motion of the trapped particle. In addition, we are able to track the Brownian motion of trapped 1 and 3 $\mu$m diameter beads from the phase jitter and show that, similar to intensity-based measurements, phase measurements can also be used to simultaneously determine displacements of the trapped bead as well as the spring constant of the trap. For lateral displacements, we have matched our experimental results with a simulation of the overall phase contour of the back-scattered light for lateral displacements by using plane wave decomposition in conjunction with Mie scattering theory. The position resolution is limited by path drifts of the interferometer which we have presently reduced to obtain a displacement resolution of around 2 nm for 1.1 $\mu$m diameter probes by locking the interferometer to a frequency stabilized diode laser. 
\end{abstract}


%

\section{Introduction}
\noindent A single microparticle trapped by optical tweezers can be used as a micro-probe which can be used in diverse applications. In some of these applications, the object of interest (such as DNA/RNA, molecular motors such as myosin and kinesin, single bacteria, etc.) is attached to the micro-probe which is held controllably in a single trap (or in some cases multiple probes trapped separately), and its motion carefully monitored to yield information about the dynamics of the object of interest including micro-forces and torques \cite{svo93, mehta99, smith01, wen07, Ghis94, volpe06}, while in other cases, the probe itself is used to reveal interesting information about surface topographies with nm precision \cite{PFM}, Brownian motion \cite{Den07, Gio07}, and to study fundamental statistical physics phenomena \cite{Mcc99}. The probe motion is manifested in changes in the forward and backward scattering patterns of a detection laser incident on the trapped probe - the scattering patterns being typically imaged on position sensitive detectors (PSD) or quadrant photodiodes (QPD). Backscattered detection is often preferred over forward scattering due to its relative independence of the morphology of the trapping apparatus, and has been shown to resolve displacements even in the pm regime \cite{Tperkins}. However, the high sensitivity obtained in displacement sensing using PSDs or QPDs is for probe motion in the radial direction, i.e. the in the direction perpendicular to the trapping beam. For motion in the axial direction, the resolution is much lower since in this case, the intrinsic resolution of QPD or PSD is no longer obtained, the motion being determined by measuring the change in the {\it total} amount of light incident on all the four quadrants of the QPD (or all the pixels of a PSD), and not the {\it difference} between pairs of quadrants (pixels for a PSD). 

In this paper, we experimentally demonstrate a technique where the {\it phase} of the back-reflected beam is measured as a function of probe motion instead of the intensity. It has been shown earlier \cite{Sin00} that phase measurement of the interference pattern formed by superposition of the backscattered light from the probe and the glass trapping chamber can be used to quantify probe axial motion. We improve the sensitivity of axial motion detection by improving the signal to noise of the interference signal by using balanced detection in an external Mach-Zender interferometer configuration. In addition, we show that even the radial motion of the probe causes a change in the phase of the interference signal, which can thus be used to calibrate radial motion as well. To calibrate the change of phase with radial motion, we perform a simulation using plane wave decomposition in conjunction with general Mie scattering theory to find out the phase shift as a function of probe displacement in the radial direction. Our method is also sensitive to Brownian motion of the trapped probe - which we are able to quantify for a given trap stiffness from the phase jitter of the interference pattern. Also, as is the case in intensity-based detection, a fourier transform of the phase jitter yields the corner frequency of the probe, and thus the spring constant or stiffness of the optical trap \cite{berg}. 

\section{Materials and methods}
\subsection{Experimental apparatus and methods}
Our optical tweezers apparatus is developed around a Zeiss Axiovert A1 inverted fluorescence microscope as shown in Fig.~\ref{setup}. A 100X, 1.4 N.A. oil immersion microscope objective (Zeiss plano-apochromat, infinity corrected) is used to couple the beams into the sample chamber. Polystyrene beads of size 1.1 or 3 $\mu$m (which serve as our micro-probes) immersed in water are trapped using a focused single transverse mode 1064 nm solid state laser (Lasever LSR1064ML) beam of 600 mW power and specified M$^2$ quality factor of 1.2.  An acousto-optic modulator (AOM) is inserted in the back focal plane of the trap to move the position of the probe radially in a controlled manner. A pair of convex lenses are used to image the plane of the AOM on to the back aperture of the microscope objective to ensure that any angular deflection at the AOM gets directly mapped to an angular deflection at the back aperture of the objective without the beam walking off. The efficiency of the AOM is around 50\% for the first order, so that the maximum power available for trapping is around 180 mW after other losses in the coupling optics. Detection is performed using a separate laser beam (Toptica DL 100, wavelength 780 nm, max power 75 mW) whose power is kept low (around 1 mW finally) enough so as to not modify the trapping potential. The trapping and detection beams are combined at the input of the microscope using a polarizing beam splitter. The detection laser beam is back-reflected from the probe and collected at the microscope side port using a dichroic mirror which has very low transmission (but very high reflection) at 1064 nm but about 50\% transmission at 780 nm. Typically, this back-reflected light is incident on a miniature quadrant photo detector \cite{Sam11} to quantify the radial motion of the probe, and determine the power spectrum of the probe motion for trap calibration. The sample chamber consists of a glass slide coated with gold (top surface of the chamber) and a cover slip of 160 $\mu$m thickness (bottom surface facing the objective), with the  polystyrene probe suspension sandwiched between. The gold coated slide is used to enhance the back-reflection from the surface which is required to compensate for the low reflectance of glass  given that the power of the detection laser is quite low.  About 25 $\mu$l of the sample is used at a dilution of 1:10000. 

\begin{figure}[!h!t]
    \centering
  {\includegraphics[scale=0.4]{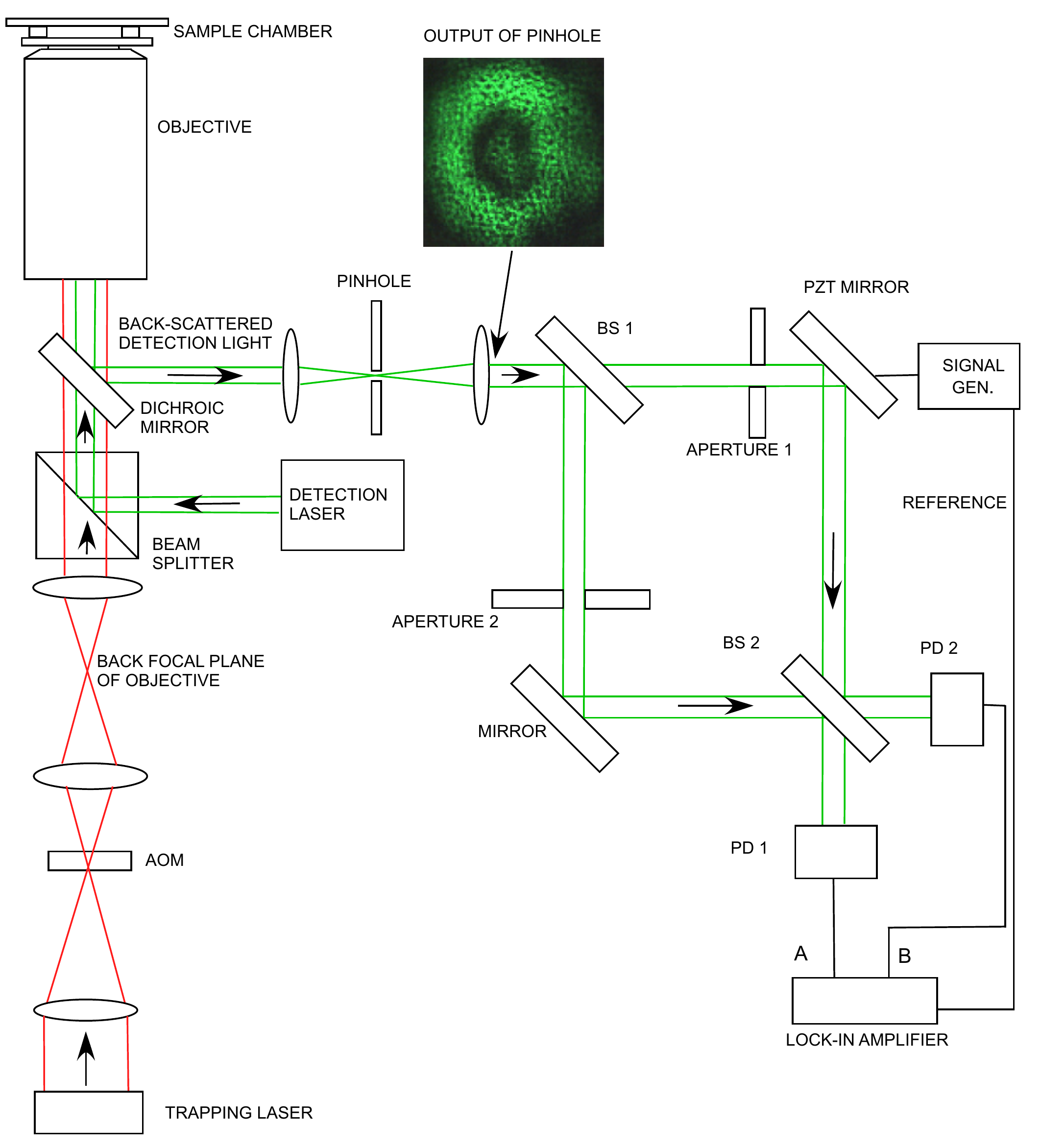}}\label{setup}
        \caption{Schematic of the experiment. Key: PD1: Photodiode 1, PD2: Photodiode 2, BS1 : Beam splitter 1, BS2: Beam splitter 2, Signal Gen.: Signal Generator. The pinhole generates an Airy pattern - the central portion of which contains scattered light from the trapped probe, while the diffuse ring contains unscattered reflection from the top gold slide of the sample chamber. Aperture 1 picks off the probe signal while Aperture 2 picks off a portion from the ring. The trapping laser is at 1064 nm and the detection laser is at 780 nm.}
 \label{setup}
   \end{figure}

The light at the microscope back-focal plane consists of back-scattered light (at 780 nm) from the trapped probe and unscattered light from different regions of the sample chamber. However, in our experiments, we trap the bead close to the top slide and separate out the scattered and unscattered components by a confocal arrangement consisting of a spatial filter used in combination with two apertures (Fig.~\ref{setup}). The spatial filter comprises of a focusing lens and a pinhole of diameter 10 $\mu$m. The focal length of the lens is carefully chosen so that the waist size at the focal spot is close to the pinhole diameter. The lens focuses the backscattered intensity pattern on the pinhole to produce an Airy pattern at the pinhole output. The diameter of the detection laser is also kept slightly larger than the probe so that the edges of the beam pass by the probe and are reflected by the top slide directly. The central region of the Airy pattern thus corresponds to light backscattered from the trapped probe (which is close to the focus of the detection laser), while the comparatively diffuse outer rings correspond to the reflected light from the gold surface. Then, using two more apertures as shown in Fig.~\ref{setup}, a portion of the light from the outer rings can be separated from the central pattern. We proceed to beat the two separated components in a Mach-Zender interferometer where the path length of one of the arms is modulated by a piezo mirror to obtain interference fringes on both photodiodes PD1 and PD2 (Thorlabs DET110). The signals on the photodiodes are out of phase by 180${\rm ^o}$, so the difference gives twice the individual signal amplitude. This is well known as balanced detection and improves our fringe contrast, and thus the sensitivity of the phase measurement by a factor of two over standard backscattered interferometry.  The difference signal is obtained using a lock-in amplifier (SR630), which also measures the phase of the output signal with reference to the phase of the driving signal to the piezo mirror. The final output of the lock-in is shown in Fig.~\ref{phfringes}(a). Now, the phase of the output signal from the interferometer changes when the probe moves with respect to the top slide, with a 2$\pi$ phase change for axial probe displacement of a unit wavelength of the detection laser.   However, the estimation of phase change for axial motion of the trapped probe with respect to the top slide needed a calibration of the axial distance moved. This could not be performed in our apparatus due to the lack of a three dimensional piezoelectric motional stage for our microscope. The probe was instead moved transversely (i.e. in the radial direction) by the AOM, and we used the fact that there is a phase change even in this case since the axial depth actually varies due to the curvature of the bead. Fringes obtained in the interferometer due to transverse motion of the bead are shown in Fig.~\ref{phfringes}(b). Experiments were performed with probes of diameter 1.1 and 3 $\mu$m. For the distance calibration, we used the pixel-to-distance calibration tool in our microscope camera (Axiovision) software and used the diameter of the probes as reference. Phase measurements as a change of probe position are shown in Fig.~\ref{findata}. Each data point was calculated by averaging over 10 independent measurements.
\begin{figure}[h]
    \centering
{\includegraphics[scale=0.35]{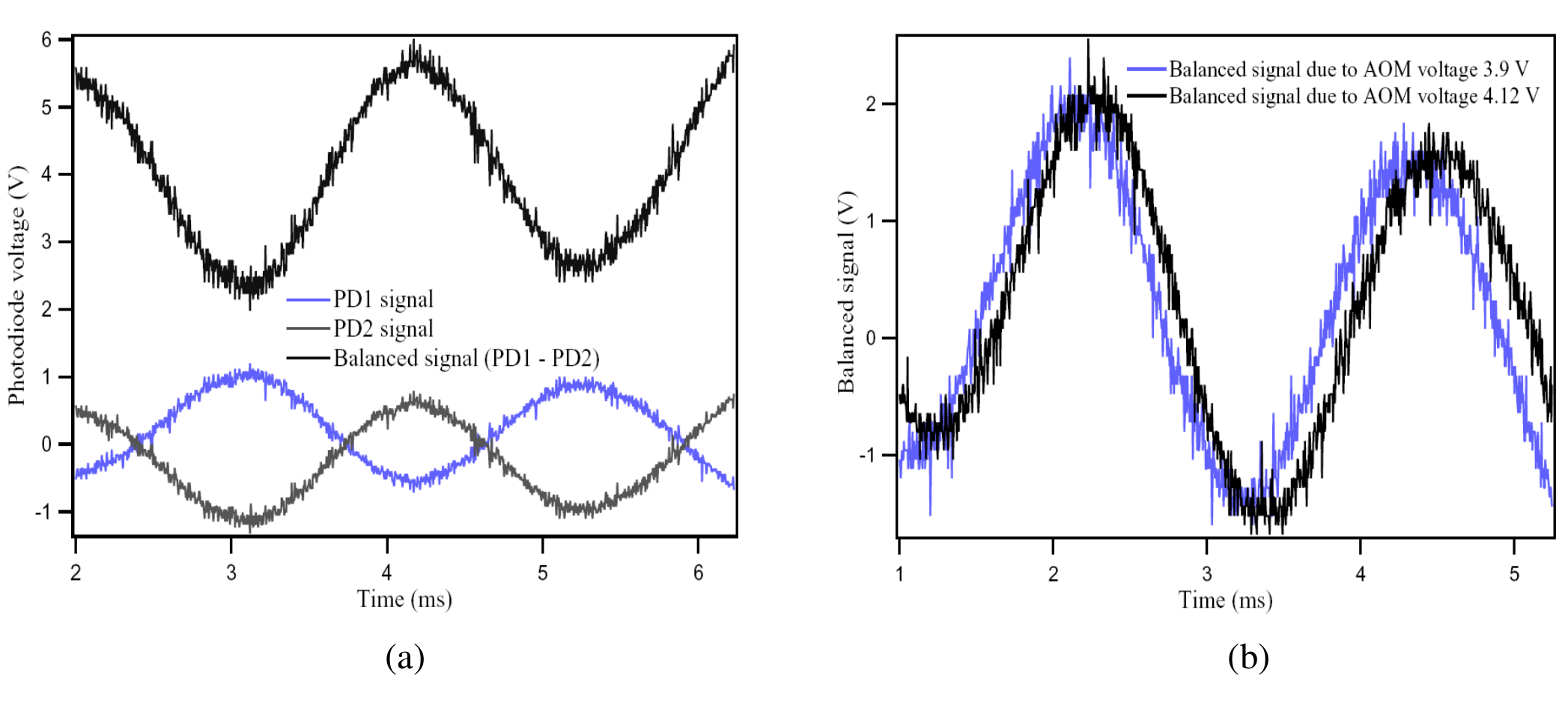}}\label{phfringes}
        \caption{(a). Balanced detection signal obtained by subtraction of the two out-of-phase signals from PD1 and PD2, the photodiodes kept in the two arms of the Mach-Zender interferometer. (b) Typical fringes obtained in the phase measurement as the AOM voltage is changed to displace the trap transversely and cause a phase shift in the backscattered signal that is captured in the balanced detection output.}
 \label{phfringes}
   \end{figure}

\subsection{Theoretical simulation}
The phase change of the backscattered signal due to axial motion is quite straight-forward (units of 2$\pi$ per half wavelength change of axial distance). On the other hand, the same change when the probe moves radially is non-trivial. A theoretical estimate was thus required to match the change in the phase of the back-scattered pattern with experimental results. While there exists literature on the intensity of the backscattered light field in optical tweezers\cite{Volpe07}, we are not aware of any study of the {\it phase} profile of the backscattered pattern. Furthermore, even Ref.~\cite{Volpe07} does not take into account the effect of the back-reflection from the top slide of a trapping chamber in calculating the intensity pattern. This is even more critical in determination of the phase profile, since the back-reflected light would interfere with the light scattered from the bead itself, thus modifying the phase profile of the overall backscattered signal. Therefore, a model was required to enable the understanding of the alterations in phase contour as the trapped microsphere was moved transversely. To develop the model, we used a variant of the Angular Spectrum Method (also referred to as vectorial Debye diffraction theory or Debye integral) \cite{Born} to calculate the electric field distribution. 

In our model, we considered a $x$-polarized Gaussian beam of light incident on the microsphere, having the form
\begin{equation}
\label{gausseq}
     E(x,y,z) = E_0 \frac{w_0}{w(z-z_w)} \exp \left( \frac{-r^2}{w^2(z-z_w)}\right) \exp \left( -ikz -ik \frac{r^2}{2R(z-z_w)} +i \zeta(z-z_w) \right)\ \hat{i}
\end{equation}
where
$r=\sqrt{x^2+y^2}$ is the radial distance, $E_0$ is the peak intensity which was set to unity, $w_0$ is the size of the waist or the tightest spot of the beam,  $z_w$ is the position of the waist relative to the $z=0$ plane, $w(z)$ is the size of the waist of the beam in the specified $z$ plane, with $w(0)=w_0$, $R(z)$ is the radius of curvature of the phase front in the specified $z$ plane, and $\zeta(z)$ is the Gouy phase shift. 

As is well known, a Gaussian beam can be decomposed into an infinite number of plane waves with appropriate weight factors. Each of these plane waves interact with the particle in accordance with the theory of Mie scattering (we consider the case when the wavelength of the light is comparable with the size of the scatterer). This decomposition was performed by applying a two dimensional fast fourier transform on the incident light field at the position of the microsphere. 
\begin{figure}[!h!t]
    \centering
\subfloat[]{\includegraphics[scale=0.3]{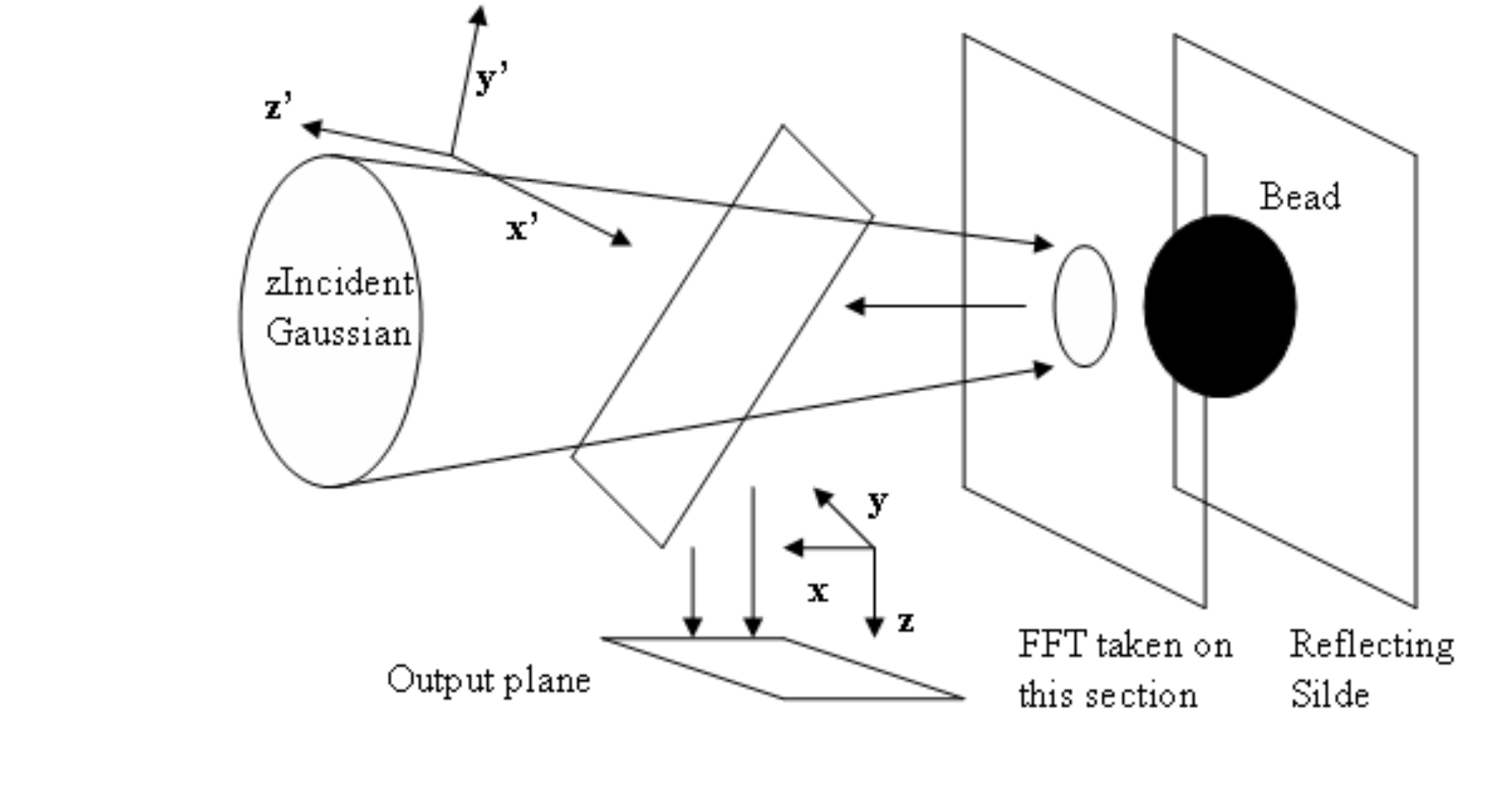}\label{simcartoon1}}
\subfloat[]{\includegraphics[scale=0.3]{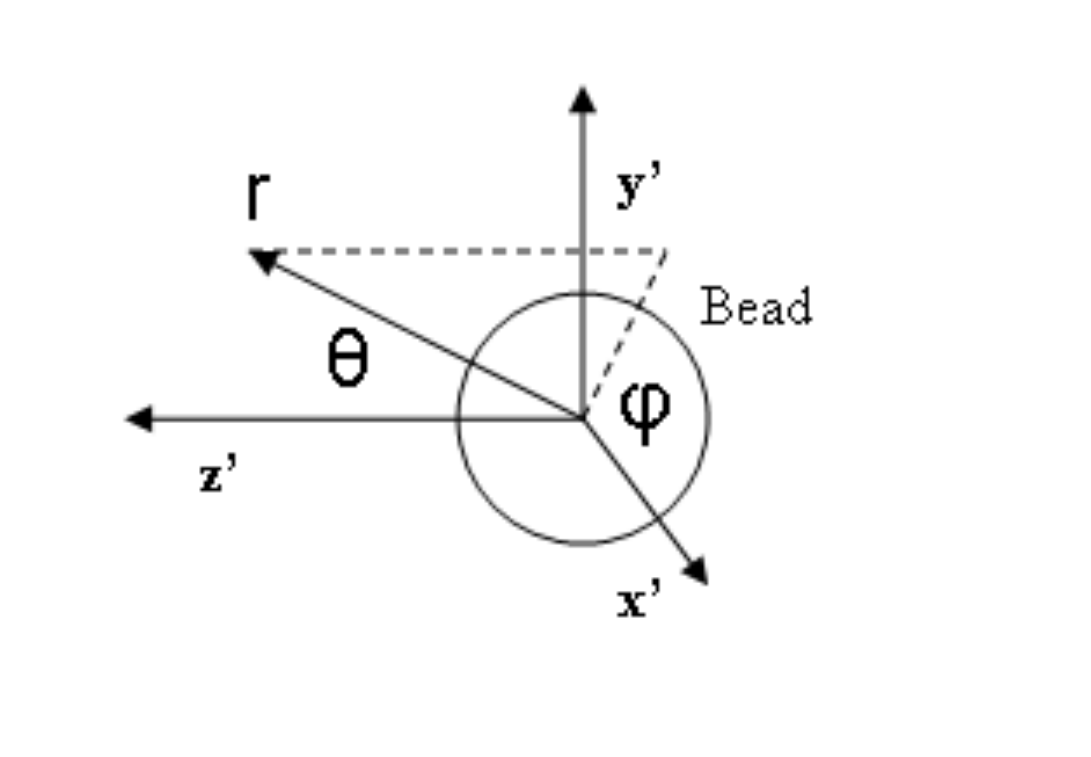}\label{simcartoon2}}
        \caption{The theoretical model}
 \label{cartoon}
   \end{figure}

\begin{equation}
E(k_x,k_y)= \sum_{n=0}^N \sum_{m=0}^M  E(x,y,z=0) e^{2 \pi i \left[k_x \left(\dfrac{n}{N}\right)+k_y \left(\dfrac{m}{M}\right)\right]}
\label{polarint1}
\end{equation}

Here, M and N are the sizes of the 2-dimensional array for the image of the cross section of the incident Gaussian beam. In each of the k-vector directions, a plane wave of unit magnitude was assumed to be generated which eventually scattered off the microsphere. The plane wave was assumed to be x polarized in it's own frame of reference (defined as $x'$). 
\begin{equation}\label{polarint2}
 E_i = e^{ikz}\bf e'_i
\end{equation}

For such an incident plane wave, the scattered wave is of the form 
\begin{equation}\label{polarint3}
 E_s = \displaystyle\sum_{n=0}^{\infty} E_n[ia_n N_{e1n}^{(3)}- b_n  M_{o1n}^{(3)} ]
\end{equation}
where, 
\begin{equation}\label{polarint4}
E_n = i^n\frac{2n+1}{n(n+1)}
\end{equation}
and $a_n, b_n$ are coefficients of scattering, while $N_{e1n}^{(3)}$ and $M_{o1n}^{(3)}$ are vector spherical harmonics with m=1 \cite{Volpe07}, 
given by the following expressions 
\begin{eqnarray}\label{polarint5}
 N_{e1n}^{(3)} &=\cos \phi\ n(n+1)\sin \theta \pi_n(\cos \theta)\dfrac{h_n^{(1)}(\varrho)}{\varrho}\bf e_r \nonumber \\
&+\cos\phi \tau_n(\cos \theta)\dfrac{\dfrac{d}{d\varrho}[\varrho h_n^{(1)}(\varrho)]}{\varrho}\bf e_\theta \nonumber \\
&-\sin\phi\ \pi_n(\cos \theta)h_n^{(1)}\dfrac{\dfrac{d}{d\varrho}[\varrho h_n^{(1)}(\varrho)]}{\varrho}\bf e_\phi
\end{eqnarray}
and
\begin{equation}\label{polarint6}
 M_{o1n}^{(3)} =\cos \phi\ \pi_n(\cos \theta)h_n^{(1)}(\varrho){\bf e_\theta} \ -\ \sin \phi \tau_n(\cos \theta)h_n^{(1)}(\varrho)\bf e_\phi
\end{equation}
In these expressions, 
\begin{center}
$\varrho = kr,$
\end{center}
with $h_n^{(1)}$ being the spherical Hankel function. 
The scattering coefficients were calculated from the boundary conditions, 
and are given by the following when the permeability of the medium and microsphere are the same.  
\begin{equation}\label{polarint7}
a_n=\frac{m\psi_n(mx)\psi_n^\prime(x)-\psi_n(x)\psi_n^\prime(mx)}{m\psi_n(mx)\xi_n^\prime(x)-\xi_n(x)\psi_n^\prime(mx)},
\end{equation}

\begin{equation}\label{polarint8}
b_n=\frac{\psi_n(mx)\psi_n^\prime(x)-m\psi_n(x)\psi_n^\prime(mx)}{\psi_n(mx)\xi_n^\prime(x)-m\xi_n(x)\psi_n^\prime(mx)},
\end{equation}

where, \begin{center}
$x=ka=\frac{2\pi n_ma}{\lambda}$,

$m=\frac{k_p}{k_m}=\frac{n_p}{n_m}$,

$\psi_n(\varrho)=\varrho j_n(\varrho)$,

$\xi_n(\varrho)=\varrho h_n^{(1)}(\varrho).$
\end{center}
Here $n_m$ is refractive index of the medium, $n_p$ is the refractive index of the microsphere and $\psi_n$, $\xi_n$ are the Ricati- Bessel functions.

The scatter patterns corresponding to each plane wave were transformed from $r, \theta$ and $\phi$ basis to the $x'$, $y'$ and $z'$ basis of the plane wave frame using the following transformation matrix 
\begin{equation}
E'_s =\left[
\begin{array}{ccc}
 sin \theta\cos \phi\ & cos \theta\cos \phi\ & sin \phi\ \\
 sin \theta\sin \phi\ & cos \theta\sin \phi\ & -cos \phi\ \\
 cos \theta\ & - sin \theta\ & 0
\end{array}
\right]
\left[
\begin{array}{ccc}
 E_r\\
 E_\theta\\
 E_\phi
\end{array}
\right]
\end{equation}

The components of the scattered field in the $x'$, $y'$ and $z'$ basis were then transformed into the x, y and z basis of the lab frame using an appropriate coordinate transformation, and eventually, scattered contributions from all the plane waves were added up to form the final backscattered pattern as shown in Eqn.~\ref{polarint9}.   
\begin{equation}\label{polarint9}
E_s =  \displaystyle\sum\limits_{k_{x'}} \displaystyle\sum\limits_{k_{y'}} E(k_{x'},k_{y'}) \left[
\begin{array}{ccc}
 cos \theta'\ cos \phi'\ & -sin \phi'\ & sin \theta'\ cos\phi'\ \\
 cos \theta'\ sin \phi'\ & cos \phi'\ & sin \theta'\ sin \phi'\ \\
 - sin \theta'\ & 0 & cos \theta'\ 
\end{array}
\right] E'_s
\end{equation}

\begin{figure}[!h!t]
    \centering
{\includegraphics[scale=0.35]{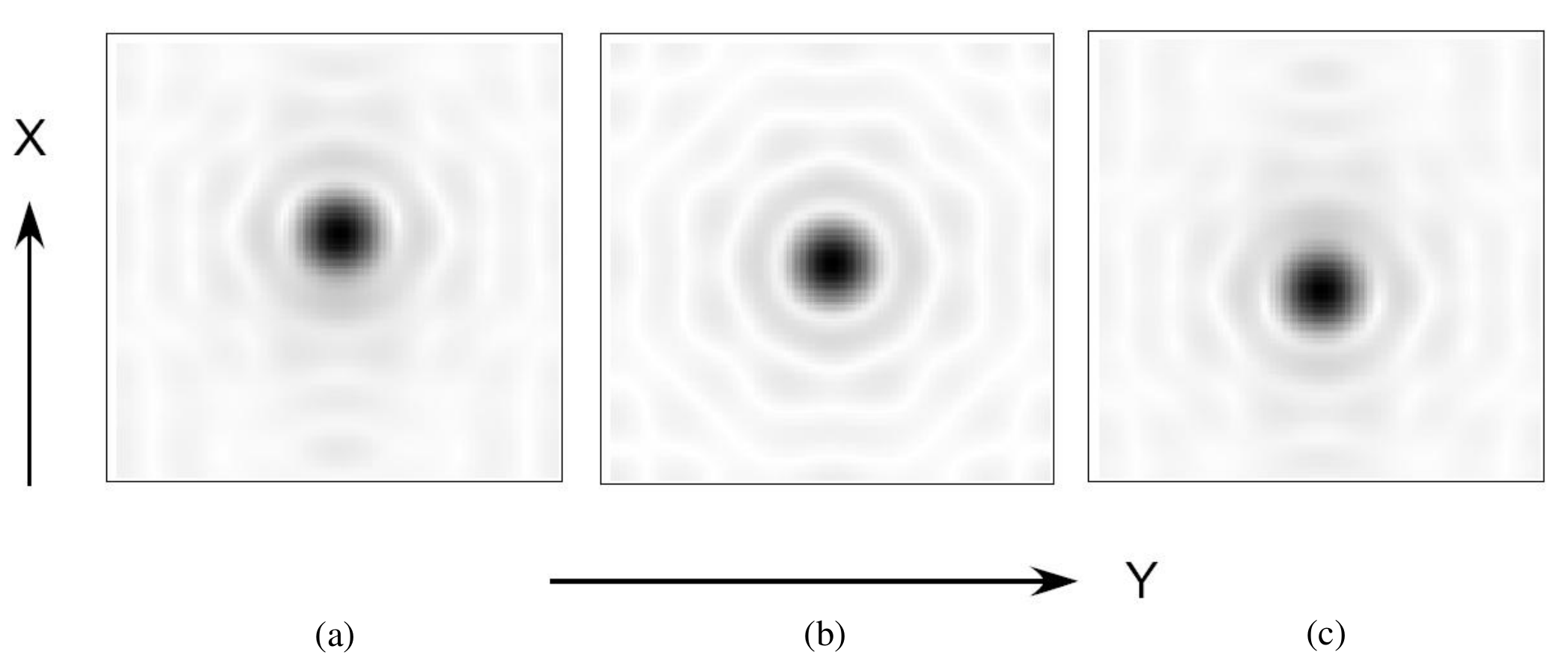}}
\caption{\label{fft}FFT of the incident Gaussian as the aperture function is offset in the x direction by (a) -1 $\mu$m (b) 0 $\mu$m (c) 1 $\mu$m. The FFT plane is assumed to be 2 $\mu$m from the focus in z direction. }
   \end{figure}

Once the back-scattered field was calculated at a certain location, the field of the light reflected from the top glass slide was also estimated at the same location by using Eq.~\ref{gausseq} and the difference in the two fields calculated. The phase contour of the resultant field was then taken. In order to simulate the shift in the longitudinal position of the microsphere from the waist of the incident detection laser beam, we used an aperture function that was moved similarly across the microsphere. The size of the aperture function was taken as the size of the microsphere itself. As the aperture function was moved, the FFT in the reciprocal plane shifted as well. The greater the offset in the radial direction, the more the shift of the FFT from the center in the reciprocal plane as demonstrated in Fig \ref{fft}. Therefore, this also produced a scatter pattern which was shifted from the center in the transverse direction. Next, the change in phase of the resultant field (superposition of scattered and reflected) was determined as a function of aperture offset by comparing the phases of the fringes at different offsets. Results of the simulation are provided in Fig \ref{findata}. 

\begin{figure}[!h!t] 
\centering
{\includegraphics[scale=0.4]{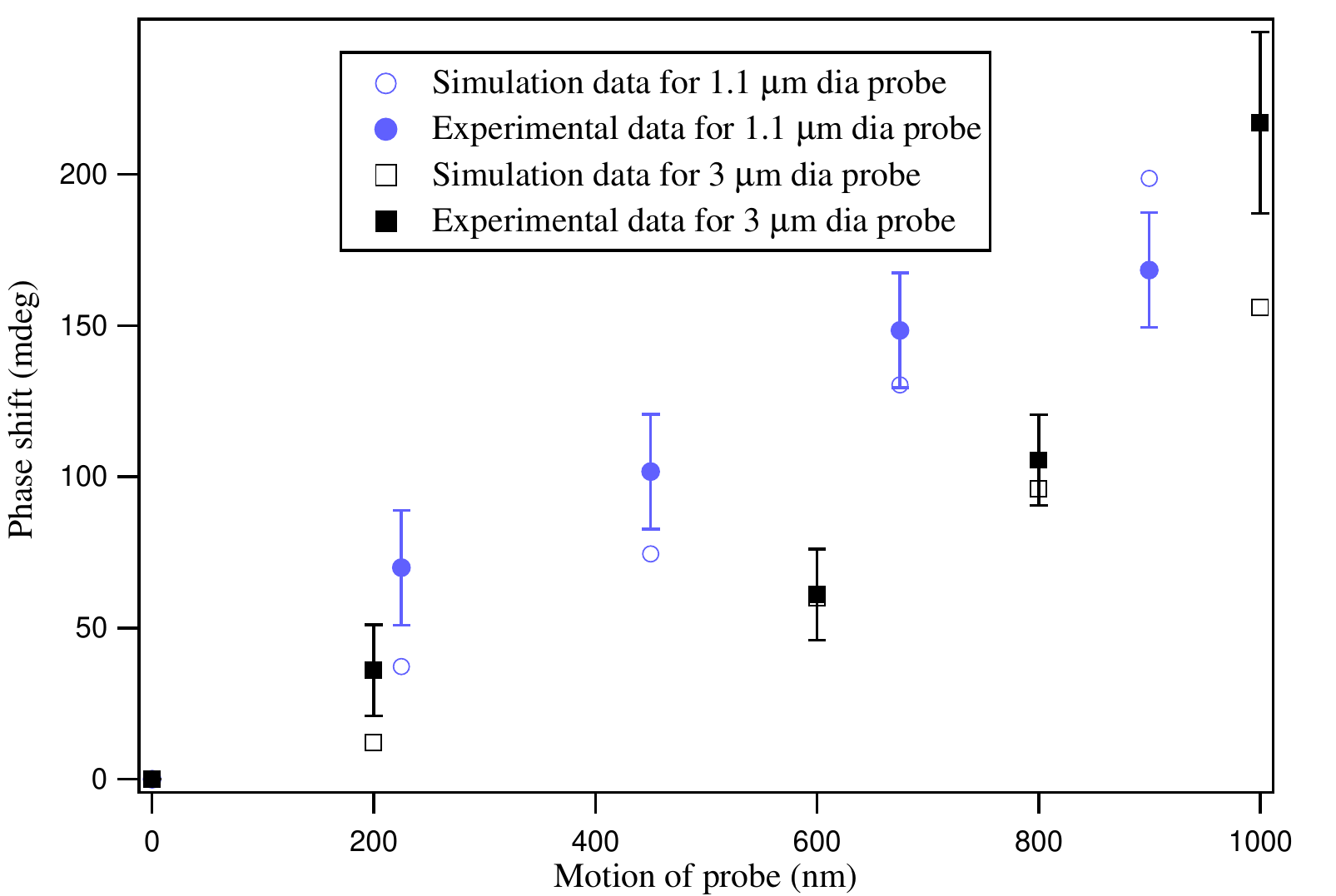}\label{findata}}
\caption{\label{findata}Experimental and simulation data for phase change for light scattered off 1.1$\mu$m and 3$\mu$m microspheres (probes) for known travel in the radial direction. In the experiment, the probes are moved in the radial direction by the AOM, while in simulation, we translate the aperture across the probe. The error bars in the experimental data signify 1$\sigma$ standard deviation. The standard deviation is predominantly due to drifts in the interferometer path length, but at low trapping powers, the Brownian motion of the trapped probe also contributes. The data point for the highest displacement of the 3 $\mu$m probe has a large error bar since the backscattered signal to noise itself was low with the bead having been displaced significantly from the detection laser.}
\end{figure}

\section{Results}
A comparison of simulation and experimental results are shown in Fig.~\ref{findata}. The experimental data points have error bars that are due to: a) path drifts of the interferometer, and b) Brownian motion of the trapped probes. The path drift is a limiting factor of our experimental technique in the sense that it determines the minimum displacement resolution achievable. In the data shown here, the path drifts were of the order of 15 degrees that corresponded to around 75 nm in terms of displacement. Note that the transfer function between displacement and phase shift was determined by a straight line fit to the theoretical data, the slope of which was $4.9 \pm 0.8$ $\mu$m/deg for 1.1 $\mu$m diameter probes. However, the path drifts can be reduced by locking the interferometer to a frequency stabilized diode laser by a simple side of fringe locking technique. By this, we managed to reduce the phase drifts to around 40 mdeg over averaging times of 100 ms, so as to give a possible displacement resolution of about 2 nm at a bandwidth of 10 Hz for 1.1 $\mu$m diameter probes. The phase measurement technique can then be implemented by either scanning the interferometer by a master-slave technique \cite{Ban04}, or by using a second frequency stabilized laser that can be scanned independently to produce fringes in the interferometer. It is interesting to note that the radial displacement resolution can be higher for smaller probes since the curvature of such probes would be higher resulting in greater phase change for smaller displacements. However, the backscattering cross section would also be lower in these cases, resulting in reduced signal to noise that would serve as a check to the resolution achievable. The interferometer path drifts could be reduced by using temperature stabilized or fiber-based cavities that could thus increase the resolution of phase measurement and even achieve sub-nm resolution in displacement sensing in axial as well as radial probe motion. 

\begin{figure}[!h!t] 
\centering
{\includegraphics[scale=0.4]{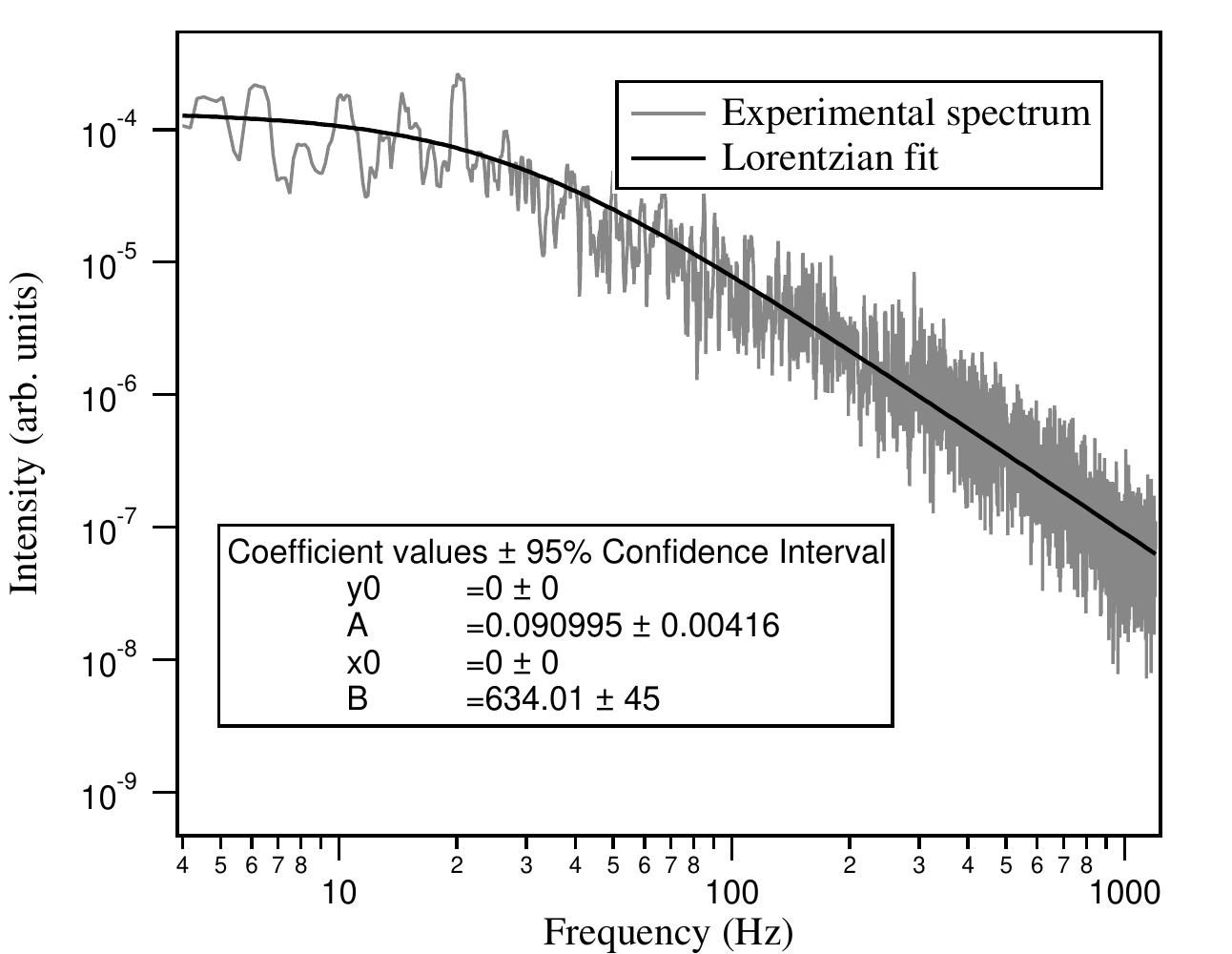}}
\caption{\label{powerspec}Power spectrum obtained by fourier transform of the phase jitter for a trapped probe of diameter 1.1 $\mu$m for an optical power of 47 mW. The data is fitted to a Lorentzian using IGOR fitting software. The fit parameters are shown, with the corner frequency given by $f_c = \sqrt(B)$ coming out to be about 25 Hz.}
\end{figure}

The Brownian motion of the trapped probes was also detected in our measurements. This was evident from the fact that the standard deviation in the data for optical powers of 47 and 120 mW comes out to be 19.0(0.3) and 16.1(0.2) deg respectively, the uncertainties being the statistical uncertainties after averaging 10 times. While the standard deviation due to path drifts is 15(0.2) deg, the enhanced standard deviation would have to be due to another independent process, which from the inverse dependence on power, could only be Brownian motion. From the standard theory of errors, we could thus use the fact that $\sigma_{tot}^2 = \sigma_{drift}^2 + \sigma_{bm}^2$, where $\sigma_{tot}^2$ is the total measured variance, $\sigma_{drift}^2 {\rm and}~ \sigma_{bm}^2$ being the variances due to path drifts and Brownian motion respectively. Since $\sigma_{tot}^2 {\rm and} ~\sigma_{drift}^2$ are known from experiment, $\sigma_{bm}^2$, and thus $\sigma_{bm}$ - the standard deviation due to Brownian motion can be determined. This comes out to be 55(5) nm and 25(5) nm for optical powers of 47 and 120 mW respectively. These results could now be compared to theoretical estimates of Brownian motion at the given trapping powers. From Ref.~\cite{Czer09}, the amplitude of Brownian motion of a trapped probe of radius $r$ is given by the expression
\begin{equation}\label{polarint1}
\delta s=\sqrt{\frac{k_B T 6 \pi \eta r}{\kappa^2 t_{ave}}},
\end{equation}
Here, $\kappa$ is the spring constant of the optical trap, $k_B$ is the Boltzmann constant, $T$ is the temperature, $\eta$ is the dynamic viscosity of water, and $t_{ave}$ is the averaging time. The spring constant can be determined by a measurement of the corner frequency $f_c$ of the trapped particle from the expression 
\begin{equation}
\label{f_c}
f_c\equiv\kappa/(2\pi\gamma_0)
\end{equation}
where, $\gamma_0 = 6\pi r \eta$. A standard measurement of corner frequency can be performed by the power spectrum method \cite{berg}. It is well known that a trapped probe executing Brownian motion obeys a simplified Langevin equation so that the power spectrum of the probe motion is a Lorentzian. Now, a position sensitive detector or a quadrant photodiode is used to record the displacement of the probe using a detection laser so that the power spectrum can then be obtained. In our technique, the phase has a linear relationship with the displacement - thus, a fourier transform of the jitter in the phase yields a similar power spectrum as is shown in Fig.~\ref{powerspec}. A Lorentzian fit to the data yields a corner frequency of around 25 Hz, which leads to a trap stiffness of 1.3 pN/$\mu$m for an optical power of 47 mW. Then, using $a = 0.55 \mu$m, an integration time of $t_{ave}$ of 8.3 ms, temperature T of 300K, viscosity $\eta$ of 0.008 Poise, one obtains the theoretical estimate of Brownian motion as 49(2) nm, the error being due to the fit error in the corner frequency. This is within 1$\sigma$ of our experimental estimate of Brownian motion at the same power. A similar measurement at an optical power of 60 mW yields a theoretical estimate of 21(2) nm, which is again within 1$\sigma$ of the experimental measurement. At higher powers, the path drifts dominated the phase jitter and it was not possible to separate out the Brownian motion of the probe. As a check of the corner frequency measurements using the phase jitter, we performed power spectrum measurements of the {\it displacement} jitter of the probe at the same power levels using our quadrant photodiode detection system \cite{Sam11}, and found an agreement to within 10\%. We can therefore conclude that our phase measurements can be used to determine probe displacements to the level of a few nm, as well as trap stiffness for an optically trapped probe in photonic force microscopy. The technique is more sensitive to axial displacement, however, we can use it to determine radial motion as well with the help of a transfer function determined from experiment and verified by simulation.   

\section{Conclusion}
In conclusion, we have developed a phase sensitive interferometric technique for simultaneously measuring probe displacements and trap parameters in a photonic force microscopy setup. The technique is based on balanced detection of the output of two arms of a Mach-Zender interferometer set up using the backscattered light from the trapped probe and its environment, and reduces the inherent lack of sensitivity in standard back focal plane interferometry for axial probe displacements. We have extended the technique to measure radial probe displacements and matched our experimental results for radial motion with a theoretical simulation. The simulation was performed using plane wave decomposition in conjunction with Mie scattering theory to find out the phase distribution of the backscattered signal due to radial probe motion, also taking into account the effects of a sample chamber. In addition, our technique is sensitive to Brownian motion, and can be used as such in any experiment to determine Brownian motion optically. The displacement sensitivity is limited mostly by the path drifts of the interferometer, that we have controlled presently to a level where we can achieve a resolution of around 2 nm for 1.1 $\mu$m diameter probes. This could be improved further by using temperature stabilized or fiber-based cavities, so that the capabilities of the technique can be extended to achieve sub-nm resolution for probe displacement in photonic force microscopy. 

\section{Acknowledgement}
This work was supported by the Indian Institute of Science Education and Research, Kolkata, an autonomous research and teaching institute funded by the Ministry of Human Resource Development, Govt of India. 


\end{document}